\documentclass{article}
\usepackage{graphicx}
\usepackage{subfigure}
\usepackage[english]{babel}
\usepackage{lineno}

\title{
\includegraphics[width=0.35\textwidth]{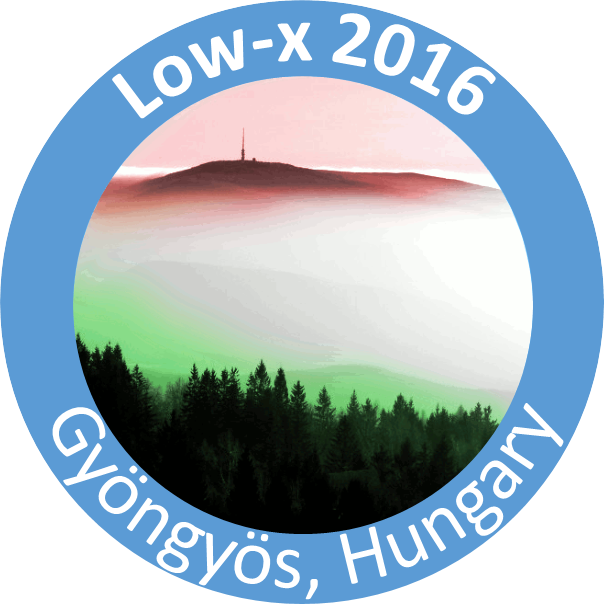}\\[1cm]
Measurements of charged-particle distributions with the ATLAS detector}
\author{{Valentina Maria Martina CAIRO$^1$$^2$, on behalf of the ATLAS Collaboration}\\[1ex]
$^1$University of Calabria, Arcavacata di Rende, Italy\\
$^2$CERN, Meyrin, Switzerland\\
}
\begin{document}

\def\pt{\ensuremath{p_\mathrm{T}}}
\def\nev{\ensuremath{N_{\mathrm{ev}}}}
\def\nch{\ensuremath{n_{\mathrm{ch}}}}
\def\nsel{\ensuremath{n_{\mathrm{sel}}}}
\def\Nch{\ensuremath{N_{\mathrm{ch}}}}
\def\meanpt{\ensuremath{\langle \pt \rangle}}
\def\nsel{\ensuremath{n_{\mathrm{sel}}}}

\fontfamily{lmss}\selectfont
\maketitle

\begin{abstract}
\noindent 
Inclusive charged-particle measurements probe the low-energy region of the non-perturbative quantum chromodynamics. The ATLAS collaboration has recently measured the charged-particle multiplicity and its dependence on transverse momentum and pseudorapidity in special data sets with low LHC beam currents, recorded at centre-of-mass energies of 8 TeV and 13 TeV. The measurements at 8 TeV cover a wide spectrum using charged-particle selections with minimum transverse momentum of both 100 MeV and 500 MeV and in various phase space regions of low and high charged-particle multiplicities, some of which are studied for the first time by ATLAS. The measurements at 13 TeV also present detailed studies with a minimum transverse momentum of both 100 MeV and 500 MeV. The measurements are compared with predictions of various tuned Monte Carlo generators and are found to provide strong constraints on these. None of the Monte Carlo generators with their respective tunes are able to reproduce all the features of the data.
\end{abstract}

\section{Introduction}
The measurements of inclusive charged-particle spectra provide insight into the low energy non-perturbative region of quantum chromodynamics (QCD). A description of low-energy processes within a perturbative framework is not possible in this regime, thus charged-particle interactions are typically described by QCD-inspired models implemented in Monte Carlo (MC) event generators. Measurements are used to constrain the free parameters of these models. 
Furthermore, soft processes, arising from pile-up at high luminosity, which leads to more than one
interaction per beam crossing, may also affect the topologies of events triggered by a specific hard-scattering interaction. 
An understanding of soft QCD processes is therefore important both in its own right and as a means of reducing systematic uncertainties in measurements of high transverse momentum phenomena.
Charged-particle distributions have been measured previously in hadronic collisions at various centre-of-mass energies, see Refs. \cite{MinBiasATLAS,MinBiasCMS1,MinBiasCMS2,MinBiasCMS3,MinBiasALICE,MinBiasCDF,MinBiasCMS4} and references therein.
This note describes the most recent charged-particle spectra measured by using data collected with the ATLAS detector \cite{ATLAS} at the centre-of-mass energy of 13 TeV~\cite{MinBias13TeV500MeV,MinBias13TeV100MeV}, with a particular emphasis on the tracking-related aspects. Some highlights from the high charged-particle multiplicity regions studied at the 8 TeV~\cite{MinBias8TeV} centre-of-mass energy are also given.
The average primary charged-particle densities at central pseudorapidity are compared to measurements at lower centre-of-mass energies.

\section{Methodology}
The methodology used in the 8 and 13 TeV analyses is similar to that used at lower centre-of-mass
energies in ATLAS \cite{MinBiasATLAS}.  The events collected correspond to minimum-bias datasets based on inelastic $pp$ interactions.  The term {\em minimum bias}
is taken to refer to trigger and event selections which are as unrestrictive as possible for the $pp$-induced final state. The data were recorded during special fills with low beam currents and reduced focusing to give a mean number of interactions per bunch crossing below 0.005.
This procedure guarantees that the contribution from pile-up in these analyses is negligible.
The measurements use tracks from primary charged-particles, corrected for detector effects to the particle level, and presented as inclusive distributions in a fiducial phase space region.
Primary charged-particles are defined as charged-particles with a mean lifetime $\tau >$ 300 ps, either directly produced in $pp$ interactions or from subsequent decays of directly produced particles with $\tau <$ 30 ps. Particles produced from decays of particles with $\tau >$ 30 ps, called
secondary  particles, are excluded. This definition differs from earlier analyses in which charged-particles with a mean lifetime 30 $<\tau <$ 300 ps were included. 
Most of these are charged strange baryons and they have been removed due to the low reconstruction efficiency of their decay products and to large variations in the predicted rates which would lead to a significant model dependence of the results presented here.

The following distributions are presented for data and compared to MC predictions:

\vspace{2mm}
\begin{center}
$ \frac{1}{\nev} \cdot \frac{\mathrm{d} \Nch}{\mathrm{d} \eta} $,   
$ \frac{1}{\nev}\cdot \frac{1}{2 \pi  p_\mathrm{T}} \cdot \frac{\mathrm{d}^2 \Nch}{\mathrm{d} \eta \mathrm{d} p_\mathrm{T}}  $,   
$\frac{1}{\nev} \cdot \frac{\mathrm{d} \nev}{\mathrm{d} \nch} $,   
$ \langle p_\mathrm{T}\rangle $ vs $\nch $,
\vspace{2mm}
\end{center}

\noindent 
where $\pt$ is the track momentum component that is transverse to the beam direction\footnote{{\scriptsize The ATLAS reference system is a Cartesian right-handed coordinate system, with
the nominal collision point at the origin. The anti-clockwise beam direction defines the positive $z$-axis, while the positive $x$-axis is defined as pointing from the collision point to the center of the
LHC ring and the positive $y$-axis  points upwards.
The azimuth angle $\phi$ is measured around the beam axis, and the polar angle $\theta$ is measured with respect to~the $z$-axis.
The pseudorapidity
is defined as $\eta =  -\ln \tan (\theta/2) $}.}, $\eta$ is the track pseudorapidity,
$\nch$ is the number of primary charged-particles in the bin relevant to the measurement,
$\nev$ is the number of selected minimum bias events, $\Nch$ is the total number of primary charged-particles in the kinematic acceptance and $\meanpt$ is the average $\pt$ for a given number of charged-particles\footnote{{\scriptsize The factor $2 \pi  p_\mathrm{T}$ in the \pt\ spectrum comes from the Lorentz invariant definition of the cross section in terms of $d^3p$.
Furthermore, the mass-less approximation is used: $y \approx \eta$. }}.
In order to make a more complete study of particle properties in minimum-bias events, results are given for different multiplicity and kinematic selections (referred to as {\em phase spaces}).
In the most inclusive phase spaces, a minimum $\nch \geq 2$ or 1 is required and the primary charged-particle must have $\eta < 2.5$ and $\pt >$ 100 MeV (referred to as {\em extended phase space}) or 500 MeV (referred to as {\em nominal phase space}) , respectively.
In the 13 TeV case, the spectra are also measured in a phase
space that is common to the ATLAS, CMS \cite{CMS} and ALICE \cite{ALICE} detectors in order to ease comparison between experiments.  For this purpose an additional requirement of $\eta < 0.8$ (referred to as {\em reduced phase space}) is made for all primary charged-particles with $\pt >$ 500 MeV and the results can be found in \cite{MinBias13TeV500MeV}.

The PYTHIA 8 \cite{Pythia} (used as a baseline), EPOS \cite{Epos} and QGSJET-II \cite{QGSJet} MC models of inclusive hadron--hadron interactions were used to generate event samples and compare their distributions to data. 
Different parameter settings in the models are used in the simulation to reproduce existing experimental data and are referred to as {\em tunes}. For PYTHIA 8, the A2 \cite{PythiaTunes} tune is based on the MSTW2008LO PDF \cite{PDF1} while the Monash \cite{Monash} underlying-event tune uses the NNPDF2.3LO PDF \cite{PDF2} and incorporates updated fragmentation parameters, as well as SPS and Tevatron data to constrain the scaling with energy. For EPOS, the LHC \cite{EposLHC} tune is used, while for QGSJET-II the default settings of the generator are applied. Detector effects are simulated using the GEANT4-based \cite{Geant4} ATLAS simulation framework \cite{ATLASSim}. The simulation also takes into account inactive and inefficient regions of the ATLAS detector.  
The resulting datasets were used to derive corrections for detector effects, evaluate systematic uncertainties and compare to the data corrected to particle level.

\section{Charged-particle measurements at 13 TeV}
\subsection{Event Selection}
\label{sec:selection}
Collision events were selected using a trigger which required one or more minimum-bias trigger
scintillators counters (MBTS) above threshold on either side of the detector.
Each event is required to contain a primary vertex, reconstructed from at least two tracks with a minimum $\pt$ of 100 MeV, as described in \cite{Vertex}. A veto is applied on additional primary vertices arising from split vertices or secondary interactions. 
A special configuration of the track reconstruction algorithms was used for this analysis to reconstruct low-momentum tracks with good efficiency and purity. 
Similar configurations were already used in Run 1, but a more robust and efficient low-$\pt$ track reconstruction program is available in Run 2 thanks to the installation of an insertable B-layer, IBL \cite{IBL}, which provides a fourth measurement point in the pixel detector.
In the {\em nominal phase space}, events are required to contain at least one selected track, passing the following criteria: $\pt >$ 500 MeV and $|\eta| <$ 2.5; at least one pixel hit and at least six SCT hits (two, four or six SCT hits for  $\pt <$ 300 MeV,  $\pt <$ 400 MeV or  $\pt >$ 400 MeV, respectively, in the case of the {\em extended phase space}), with the additional requirement of an innermost-pixel-layer hit if expected\footnote{A hit is expected if the extrapolated track crosses a known active region of a pixel module.} (if a hit in the innermost layer is not expected,
the next-to-innermost hit is required if expected); $|d_{0}^{BL} | <$ 1.5 mm, where the transverse impact parameter, $d_{0}^{BL}$, is calculated with respect to the measured beam line position; and $|z_{0}^{BL} \cdot sin\theta| <$ 1.5 mm, where $z_{0}^{BL}$is the difference between the longitudinal position of the track along the beam line at the point where $d_{0}^{BL}$ is measured and the longitudinal position of the primary vertex, and $\theta$ is the polar angle of the track. Finally, in order to remove tracks with mismeasured $\pt$ due to interactions with the material or other effects, the track-fit $\chi^{2}$ probability is required to be greater than 0.01 for tracks with $\pt <$ 10 GeV. 

Approximately 9 million events are selected, containing a total of $\sim$ 100 million reconstructed tracks. While the overall number of particles in the kinematic acceptance of the {\em extended phase space} is nearly double that in the {\em nominal phase space}, the measurements are more difficult for $\pt <$ 500 MeV, due to multiple scattering and imprecise knowledge of the material in the detector. These systematic uncertainties at low $\pt$ need therefore to be carefully evaluated.
The performance of the Inner Detector (ID) track reconstruction in the 13 TeV data and its simulation is described in Ref.
\cite{ATL-PHYS-PUB-2015-018}. Overall, good agreement between data and simulation is observed.

\subsection{Analysis strategy}

\begin{figure}[h!]
\centering
\subfigure[]{
\includegraphics[width=0.36\textwidth]{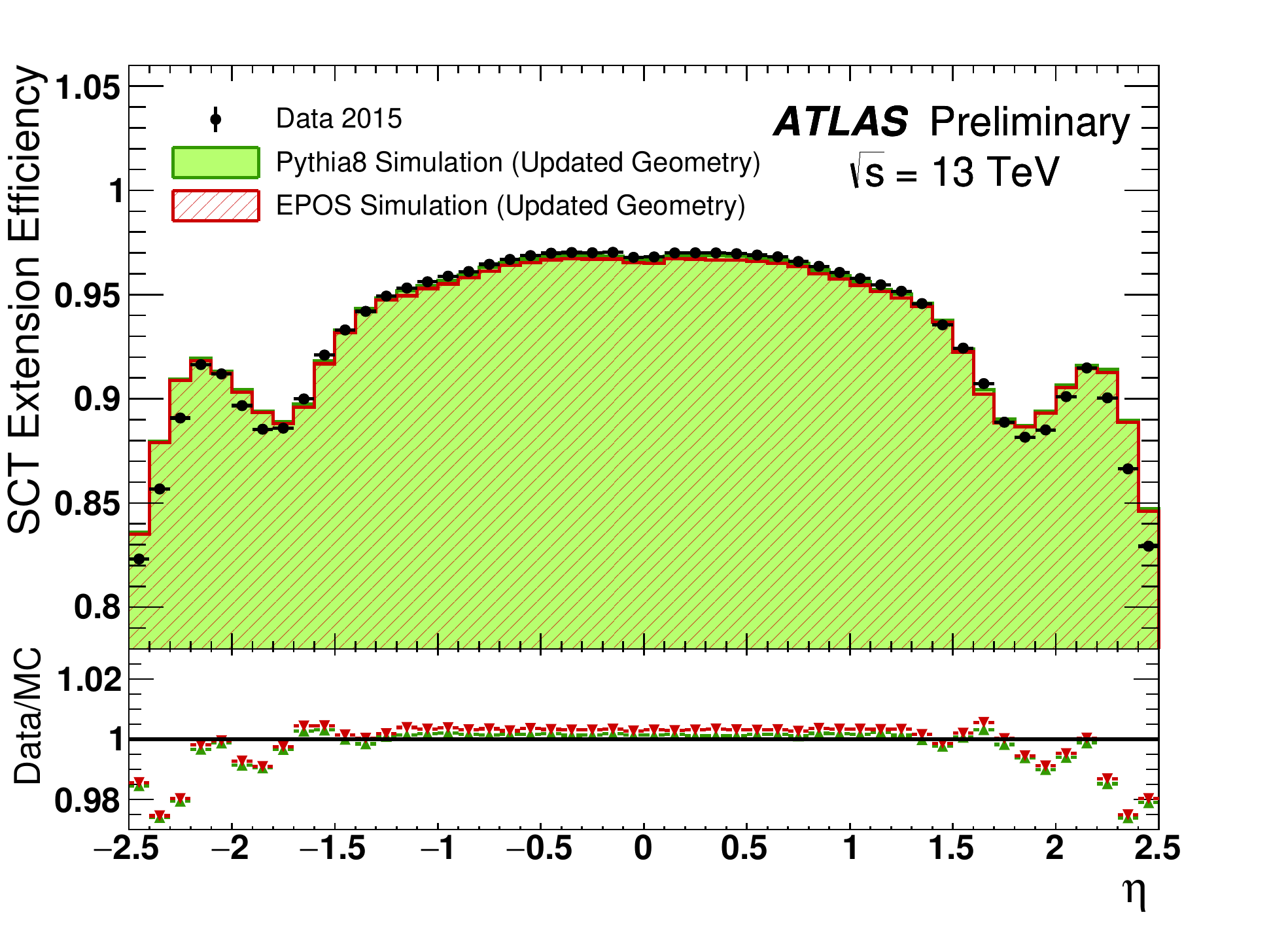}
\label{fig:SCTExtEff}
}
\subfigure[]{
\includegraphics[width=0.36\textwidth]{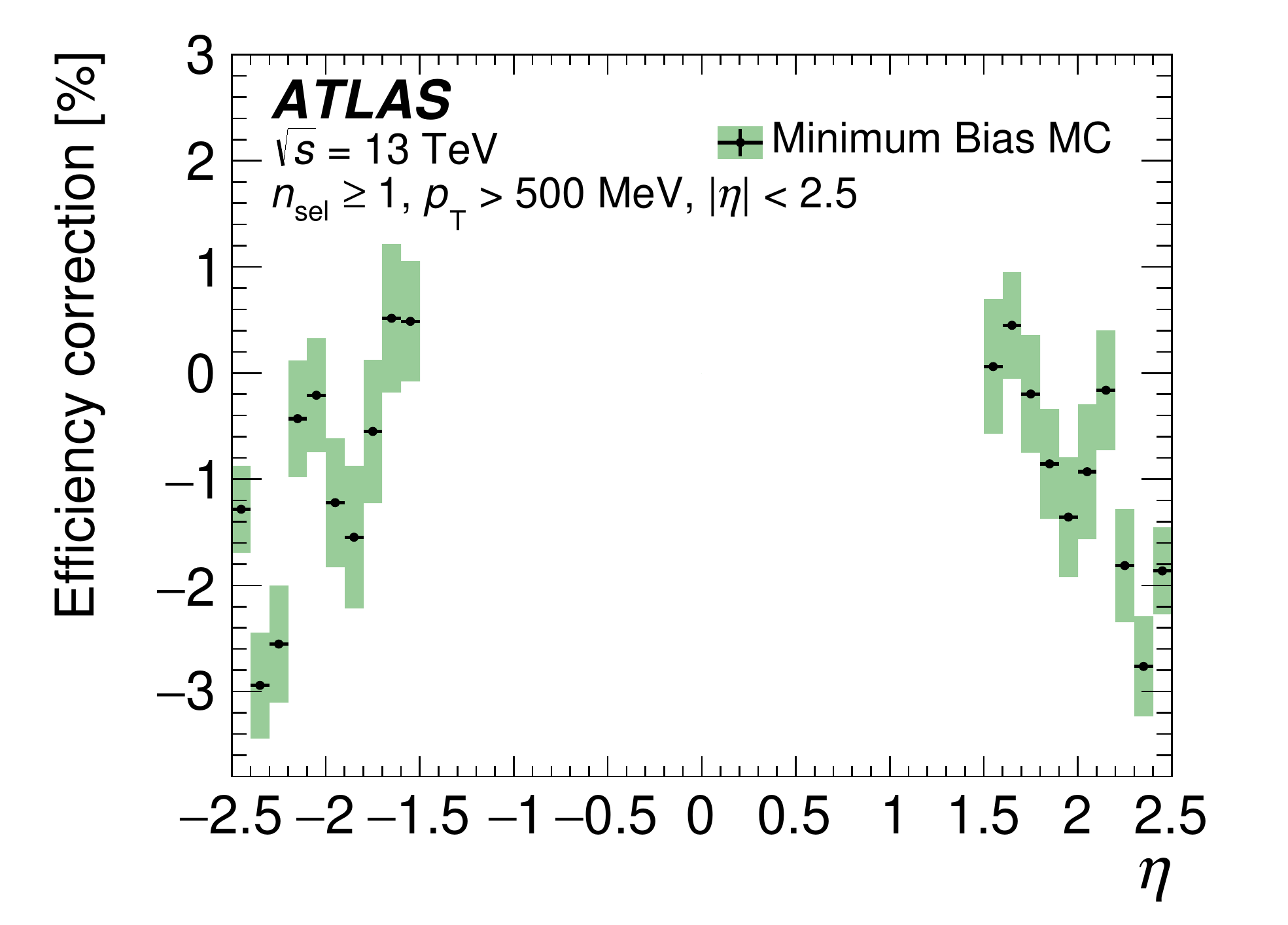}
\label{fig:correction}
}
\subfigure[]{
\includegraphics[width=0.36\textwidth]{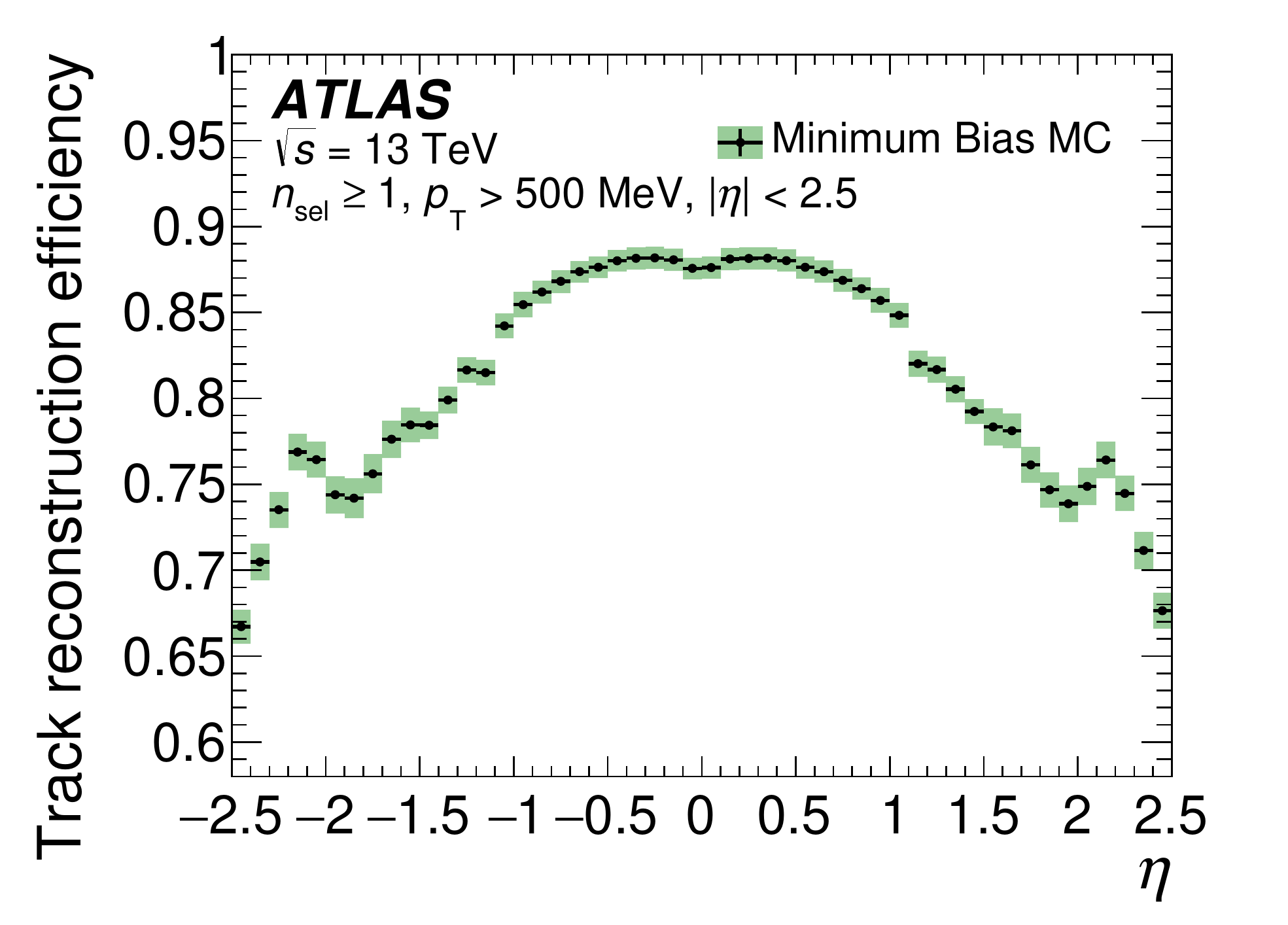}
\label{fig:effeta}
}
\caption{\subref{fig:SCTExtEff} SCT extension efficiency as a function of pseudorapidity $\eta$ of the pixel track segments in a comparison between data, PYTHIA 8 and EPOS, from Ref. \cite{Material}. \subref{fig:correction} Data-driven correction to the track reconstruction efficiency as a function of pseudorapidity, $\eta$, from Ref. \cite{MinBias13TeV500MeV}. \subref{fig:effeta} The track reconstruction efficiency after this correction as a function of $\eta$, from Ref. \cite{MinBias13TeV500MeV}. $\nsel$ is defined as the number of tracks passing all of the track selection requirements.}
\label{fig:track}
\end{figure}

The main steps of the analysis are related to the trigger, vertex and track reconstruction efficiencies, which need to be evaluated together with their uncertainties. The background contributions to the tracks from primary particles, which include fake tracks (those formed by a random combination of hits), strange baryons and secondary particles, need to be estimated as well. Observables of interest can be evaluated and, by means of an unfolding procedure, can be corrected to account for detector effects.
The details can be found in Refs. \cite{MinBias13TeV500MeV,MinBias13TeV100MeV}, while, in the next section, a few insights will be given on the track reconstruction efficiency by highlighting the importance of a precise evaluation of the amount of material in the ATLAS ID, which represents the main source of systematic uncertainty for this analysis.

\subsection{Track reconstruction efficiency}
The analysis is a track-based analysis and the evaluation of the track reconstruction efficiency and of the related systematics is crucial. The dominant uncertainty in the track reconstruction efficiency arises from imprecise knowledge of the amount of material in the ID.
The primary track reconstruction efficiency $\varepsilon _\mathrm {trk}$ is determined from simulation.
The efficiency is parameterised in two-dimensional bins of $\pt$ and $\eta$, and is defined as:
\begin{equation}
\varepsilon _\mathrm {trk}(p_\mathrm {T},\eta ) = \frac{N^\mathrm {matched}_\mathrm {rec}(p_\mathrm {T},\eta )}{N_\mathrm {gen}(p_\mathrm {T},\eta )},
\end{equation}
where $\pt$ and $\eta$ are defined at generator level, $N^\mathrm {matched}_\mathrm {rec}(\pt,\eta )$ is the number of reconstructed tracks matched to a generated primary charged-particle and $N_\mathrm {gen}(p_\mathrm {T},\eta)$ is the number of generated primary charged-particles in the kinematic region of interest. A track is matched to a generated particle if the weighted fraction of track hits originating from that particle exceeds 50\%.
In the analysis performed in the {\em nominal phase space}, a data-driven correction to the efficiency was applied in order to account for material effects in the $|\eta|>$ 1.5 region.
The track reconstruction efficiency depends on the amount of material in the detector, due to particle interactions that lead to efficiency losses. The relatively large amount of material between the pixel and SCT detectors in the region $|\eta| >$ 1.5 has changed between Run 1 and Run 2 due to the replacement of some pixel services, which are difficult to simulate accurately. The track reconstruction efficiency in this region is corrected using a method, referred to as SCT extension efficiency \cite{Material}, that compares data and simulation for the efficiency to extend a track reconstructed in the pixel detector (referred to as pixel track segment) into the SCT. Differences in SCT extension efficiency are quite sensitive to differences in the amount of material in this region, as can be seen in Figure \ref{fig:SCTExtEff}. The correction, together with the systematic uncertainty, coming predominantly from the uncertainty of the particle composition in the simulation used to make the measurement, is shown in Figure \ref{fig:correction}. The uncertainty in the track reconstruction efficiency resulting from this correction is $\pm$0.4\% in the region $|\eta |>$ 1.5. 
The resulting reconstruction efficiency as a function of $\eta$ integrated over $\pt$ is shown in Figure \ref{fig:effeta}. 
The track reconstruction efficiency is lower in the region $|\eta| >$ 1 due to particles passing through more material in that region. The slight increase in efficiency at $|\eta| \sim$ 2.2 is due to particles passing through an increasing number of layers in the end-cap Pixel regions of the ID.
The data-driven correction allows for a large reduction of the systematic uncertainty in the measurement with respect to previous studies but it cannot be applied in the {\em reduced phase space} due to the large uncertainties of this method for low-momentum tracks. In this case, the total uncertainty on the track reconstruction efficiency due to the amount of material is calculated as the linear sum of the contributions of 5\% additional material in the entire ID, 10\% additional material in the IBL and 50\% additional material in the pixel services region for $|\eta| >$ 1.5, as described in detail in \cite{Recomm}.

The SCT extension efficiency only probes the material between the pixel and SCT detectors in the region $|\eta |>$ 1.5, but a good understanding of the material in the other regions of the ID is needed for good description of the track reconstruction efficiency. The material in the ID was studied extensively during Run 1 \cite{MaterialR1, MaterialR1_bis}, where the amount of material was known to $\pm$5\%. 
This gives rise to a systematic uncertainty in the track reconstruction efficiency of $\pm$0.6\% ($\pm$1.2\%) in the most central (forward) region. 
Between Run 1 and Run 2, the IBL was installed, and its simulation therefore can only be optimised with the Run 2 data. Two data-driven methods are used \cite{Material}: a study of secondary vertices from photon conversions and a study of secondary vertices from hadronic interactions, where the radial position of the vertex and the invariant mass of the outgoing particles are measured. 
Comparisons between data and simulation indicate that the material in the IBL is constrained to within $\pm$10\%. This leads to an uncertainty in the track reconstruction efficiency of $\pm$0.1\% ($\pm$0.2\%) in the central (forward) region. 
This uncertainty is added linearly to the uncertainty from constraints from Run 1, to cover the possibility of missing material in the simulation in both cases. The resulting uncertainty is added in quadrature to the uncertainty from the data-driven correction. 

The total uncertainty on the track reconstruction efficiency due to the imperfect knowledge of the detector material is $\pm$0.7\% in the most central region and it grows to $\pm$1.5\% in the most forward region.
There is a small difference in efficiency, between data and simulation, due to the detector hit requirements described in Section \ref{sec:selection}. This difference is assigned as a further systematic uncertainty, amounting to $\pm$0.5\% for $\pt <$ 10 GeV and $\pm$0.7\% for $\pt >$ 10 GeV.
The total uncertainty due to the track reconstruction efficiency determination, shown in Figure \ref{fig:effeta}, is obtained by adding all effects in quadrature and is dominated by the uncertainty from the material description.

\subsection{Corrections and final results}
To produce unfolded distributions at particle level, all distributions are first corrected for the loss of events due to the trigger and vertex requirements. The $\eta$ and $\pt$ distributions of selected tracks are then corrected using a track-by-track weight, as described in Refs. \cite{MinBias13TeV500MeV,MinBias13TeV100MeV,MinBias8TeV}. No additional corrections are needed for the $\eta$ distribution because the resolution is smaller than the bin width. For the $\pt$ distribution, an iterative Bayesian unfolding \cite{Unf} is applied to correct the measured track $\pt$ distribution to that for primary particles.
After applying the event weight, the Bayesian unfolding is also applied to the multiplicity distribution.
The total number of events, $\nev$, used to normalise the distributions, is defined as the integral of the $n_{\mathrm {ch}}$ distribution, after all corrections are applied.
The dependence of $\langle p_\mathrm {T}\rangle$ on $n_{\mathrm {ch}}$ is obtained by first separately correcting the total number of tracks and $\sum _{i}p_{\rm{T}} (i)$ (summing over the $\pt$ of all tracks and all events), both versus the number of primary charged-particles. After applying the correction to all events using the event and track weights, both distributions are unfolded separately. The bin-by-bin ratio of the two unfolded distributions gives the dependence of $\langle p_\mathrm {T}\rangle$ on $n_{\mathrm {ch}}$.

\begin{figure}[h!]
\begin{center}
\subfigure[]{
\includegraphics[width=0.34\textwidth]{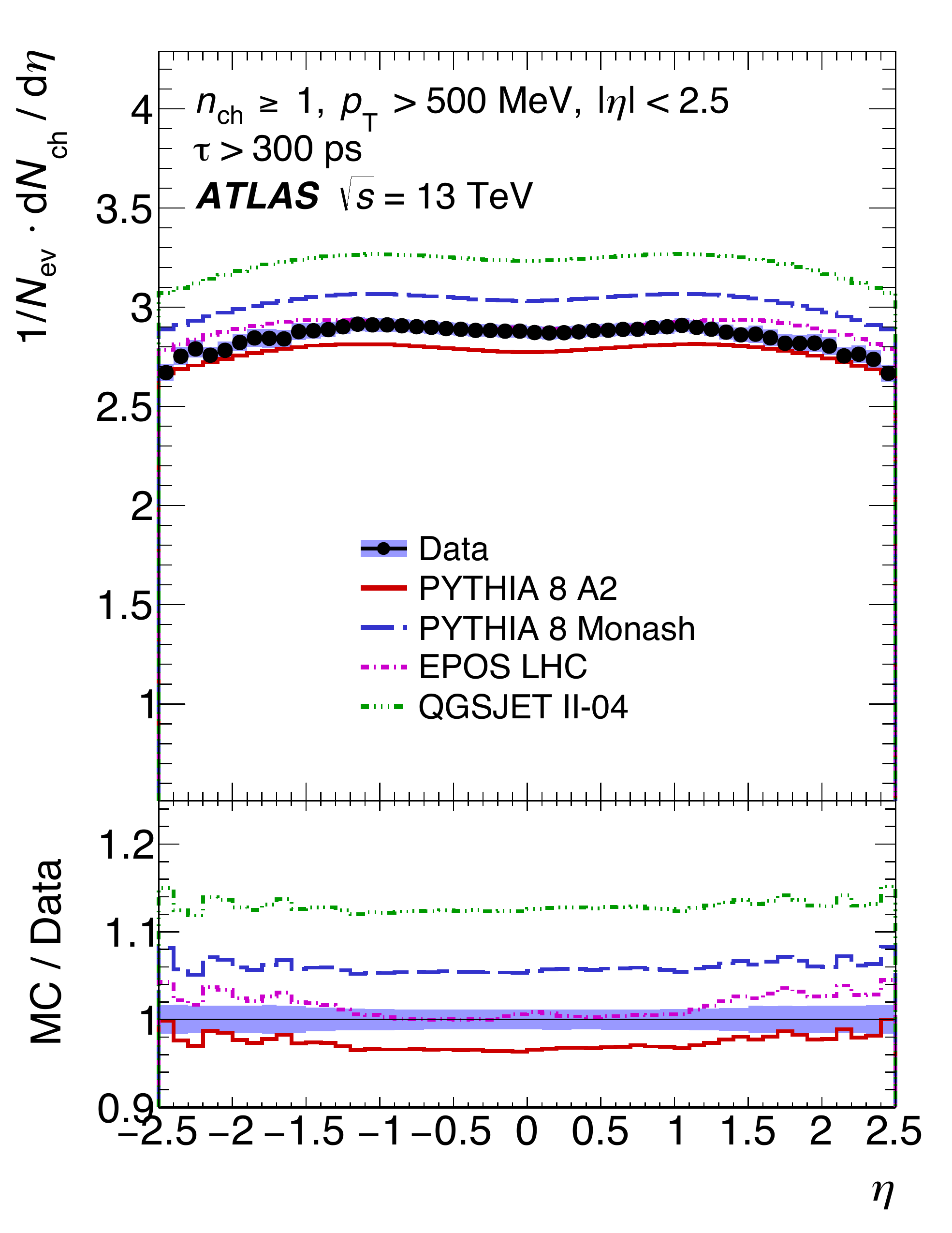}
\label{fig:500MeVEta}
}
\subfigure[]{
\includegraphics[width=0.34\textwidth]{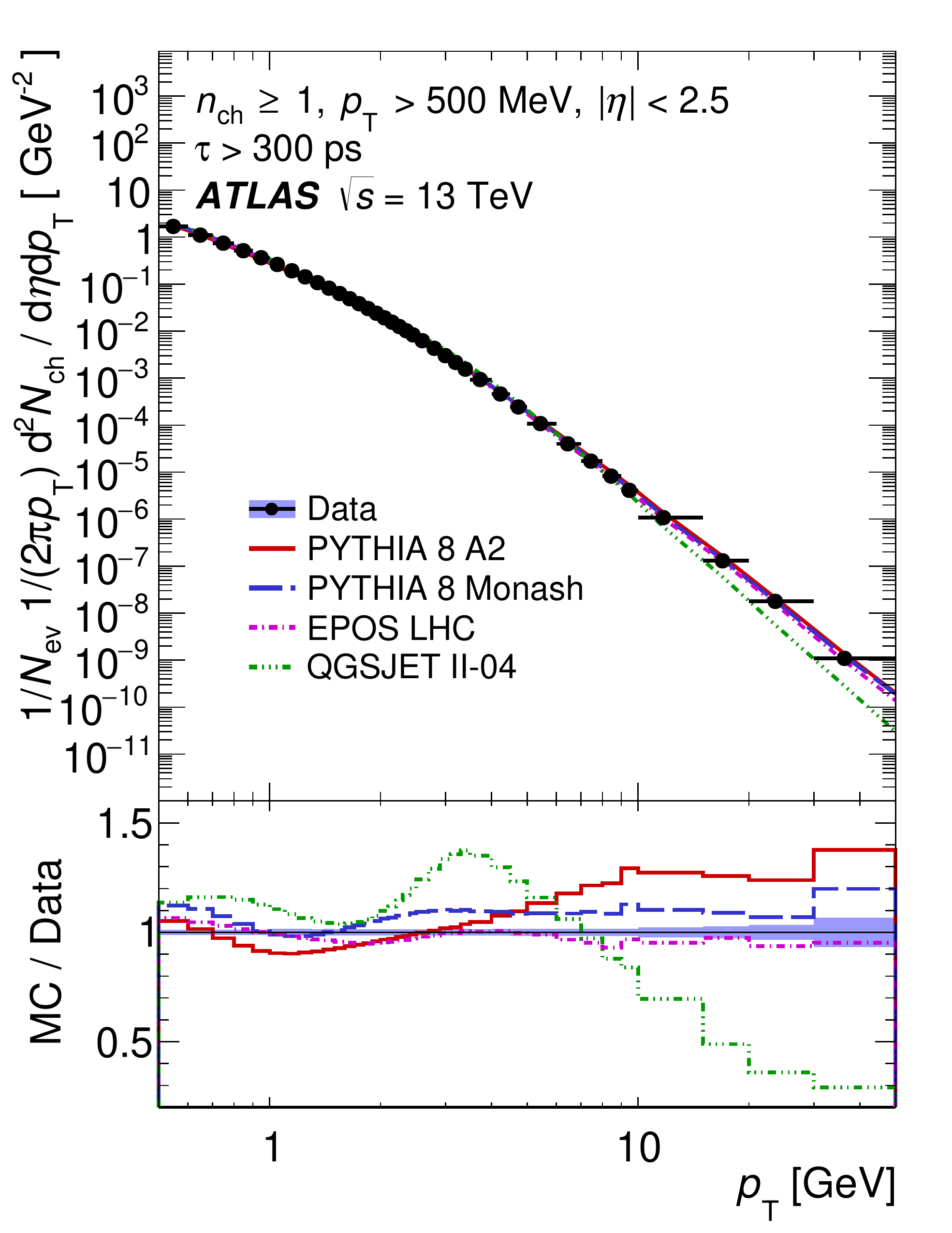}
\label{fig:500MeVPt}
}
\subfigure[]{
\includegraphics[width=0.34\textwidth]{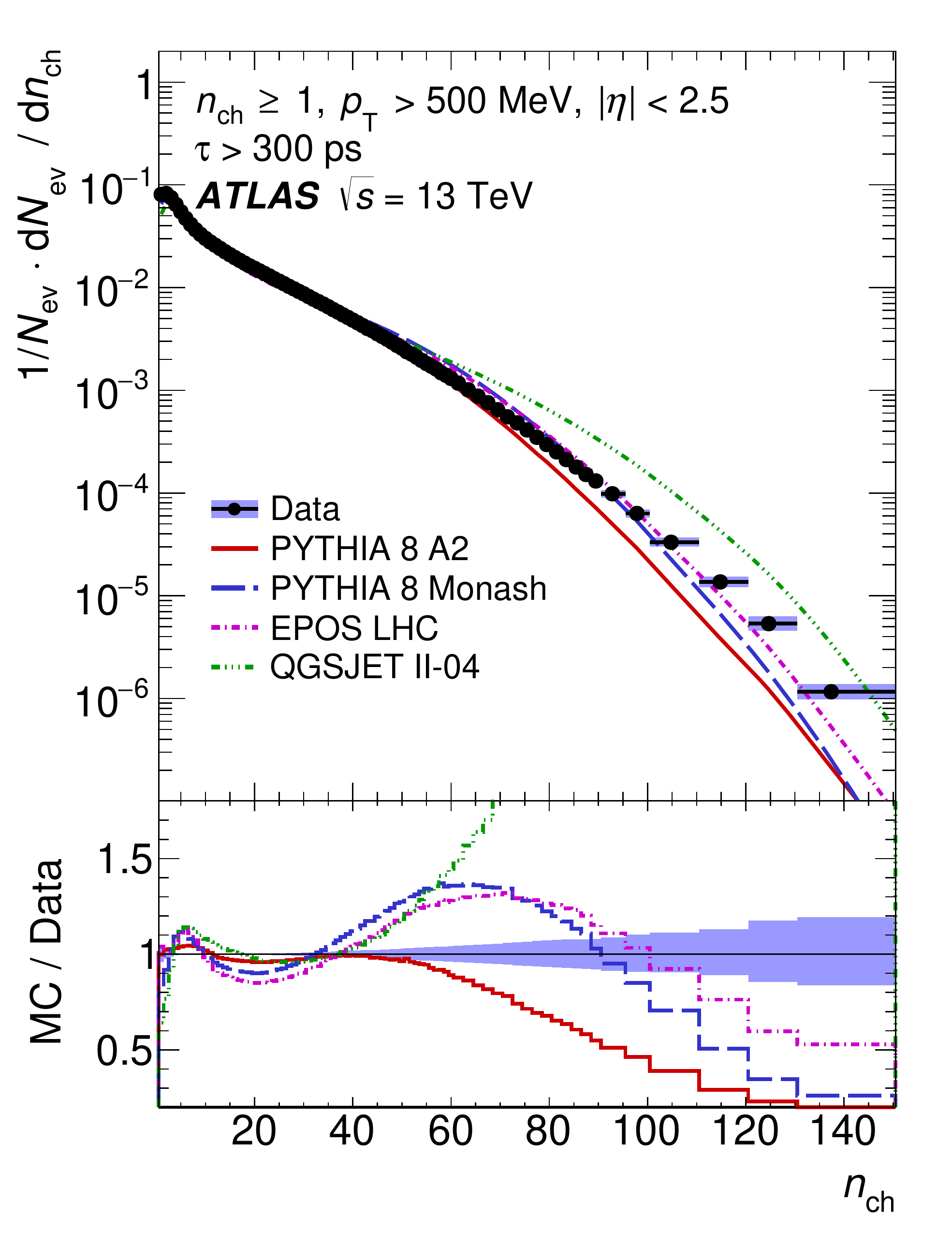}
\label{fig:500MeVnch}
}
\subfigure[]{
\includegraphics[width=0.34\textwidth]{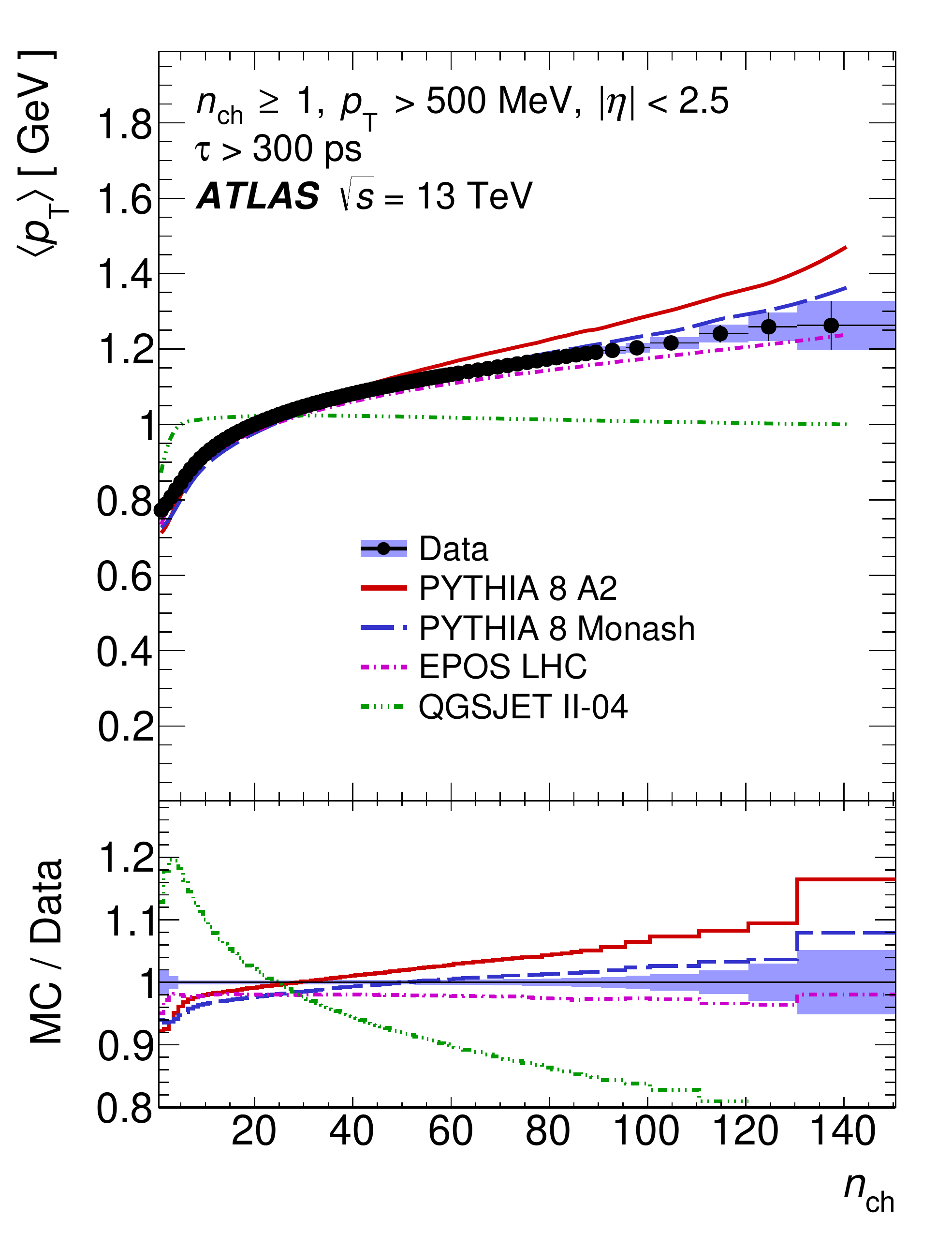}
\label{fig:500MeVptnch}
}
\caption{13 TeV data, from Ref. \cite{MinBias13TeV500MeV}: Primary charged-particle multiplicities as a function of \subref{fig:500MeVEta} pseudorapidity $\eta$ and \subref{fig:500MeVPt} transverse momentum $\pt$,  \subref{fig:500MeVnch} the primary charged-particle multiplicity $\nch$ and \subref{fig:500MeVptnch} the mean transverse momentum $\meanpt$ versus $\nch$ for events with at least one primary charged-particles with $\pt > $ 500 MeV, with $|\eta| <$ 2.5, and with a lifetime $\tau >$ 300 ps. 
}
\label{fig:Final500MeV}
\end{center}
\end{figure}

\begin{figure}[h!]
\begin{center}
\subfigure[]{
\includegraphics[width=0.34\textwidth]{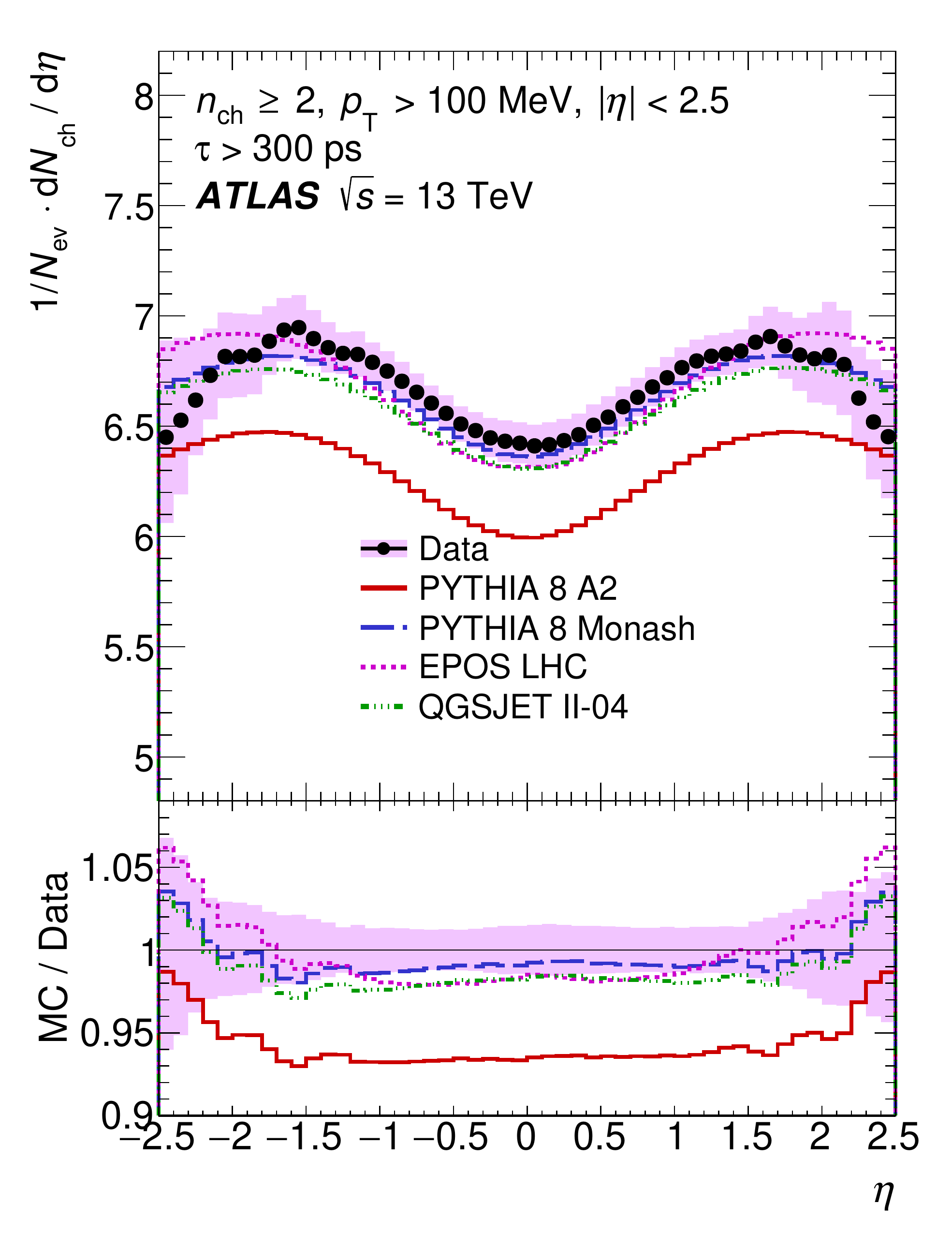}
\label{fig:100MeVEta}
}
\subfigure[]{
\includegraphics[width=0.34\textwidth]{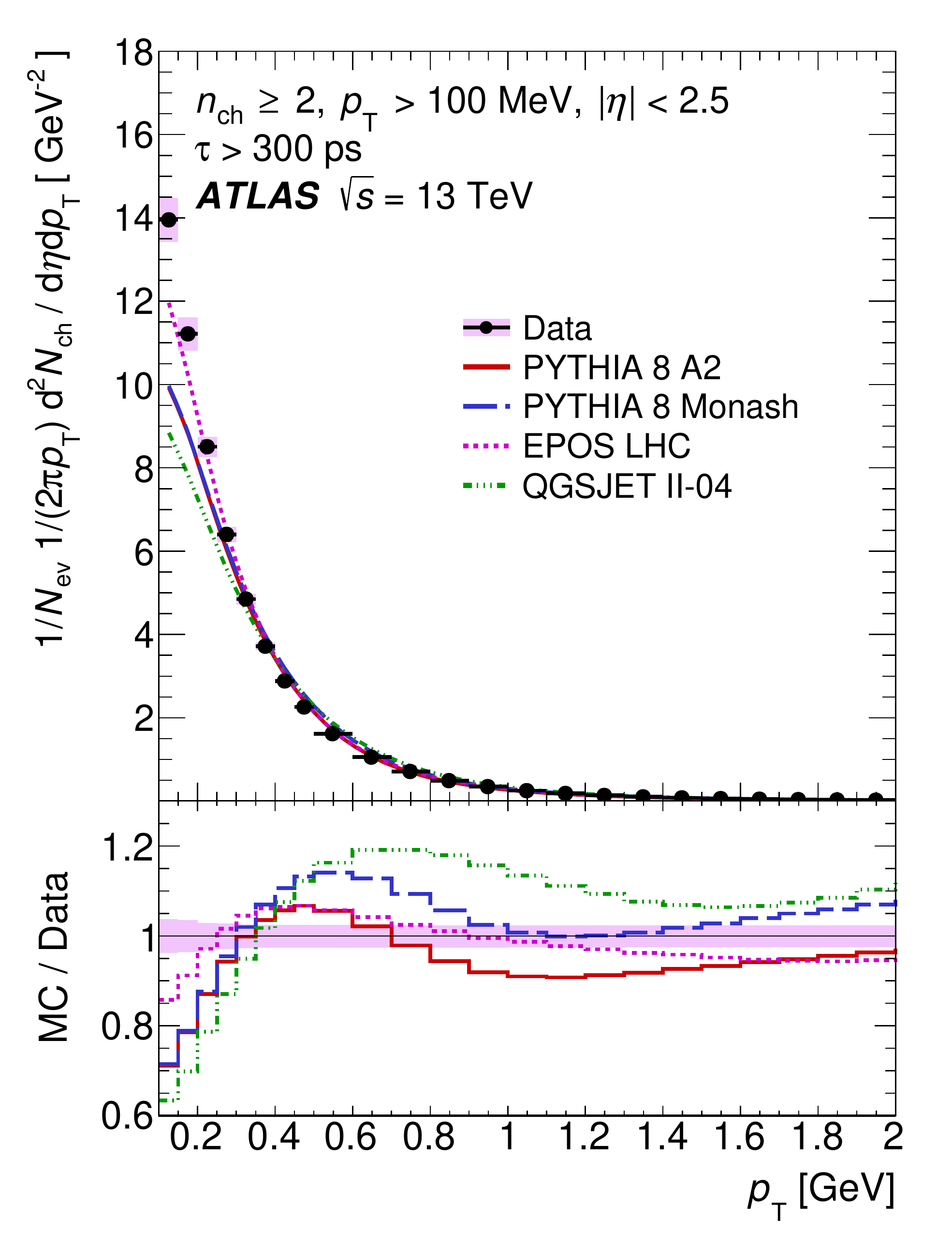}
\label{fig:100MeVPt}
}
\subfigure[]{
\includegraphics[width=0.34\textwidth]{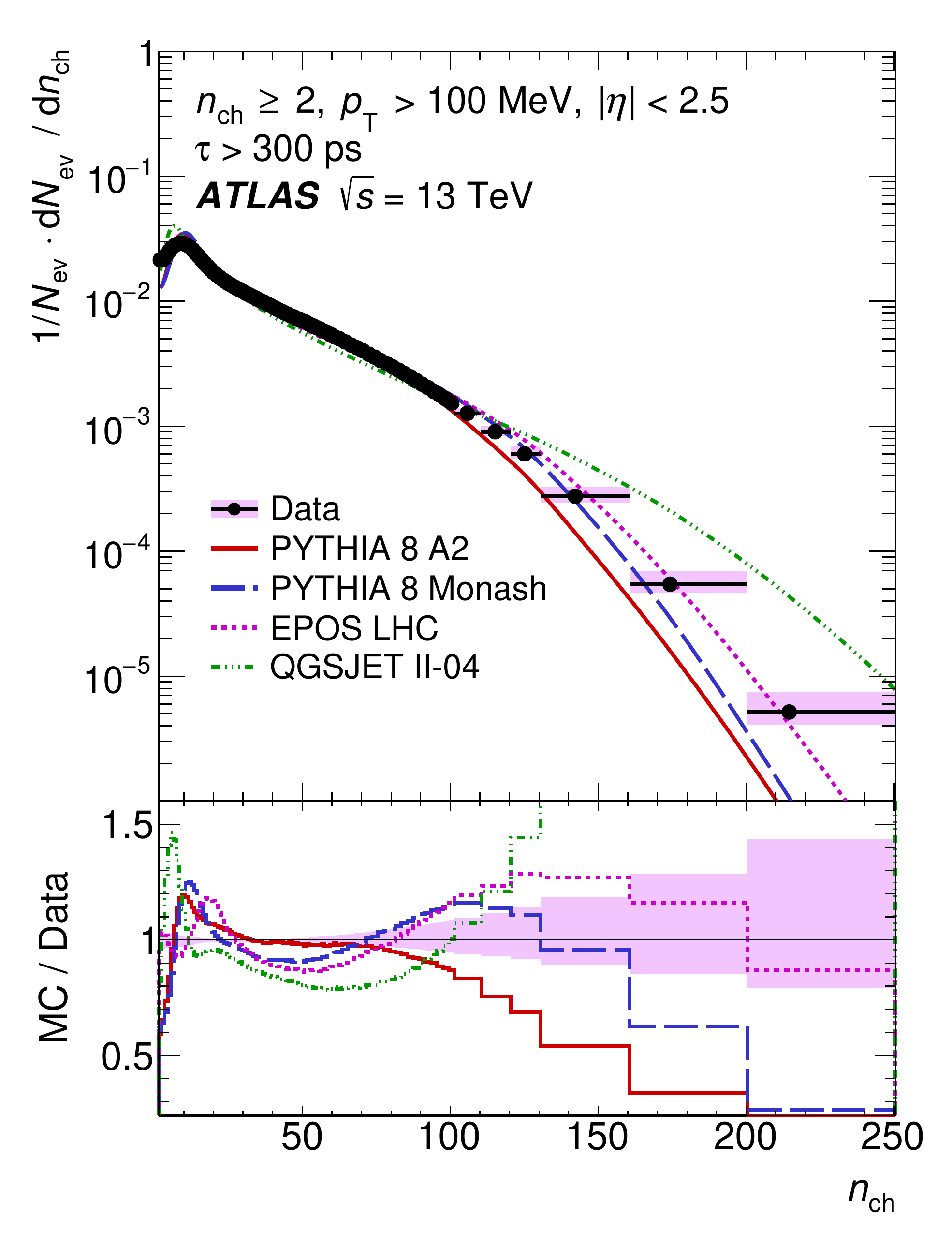}
\label{fig:100MeVnch}
}
\subfigure[]{
\includegraphics[width=0.34\textwidth]{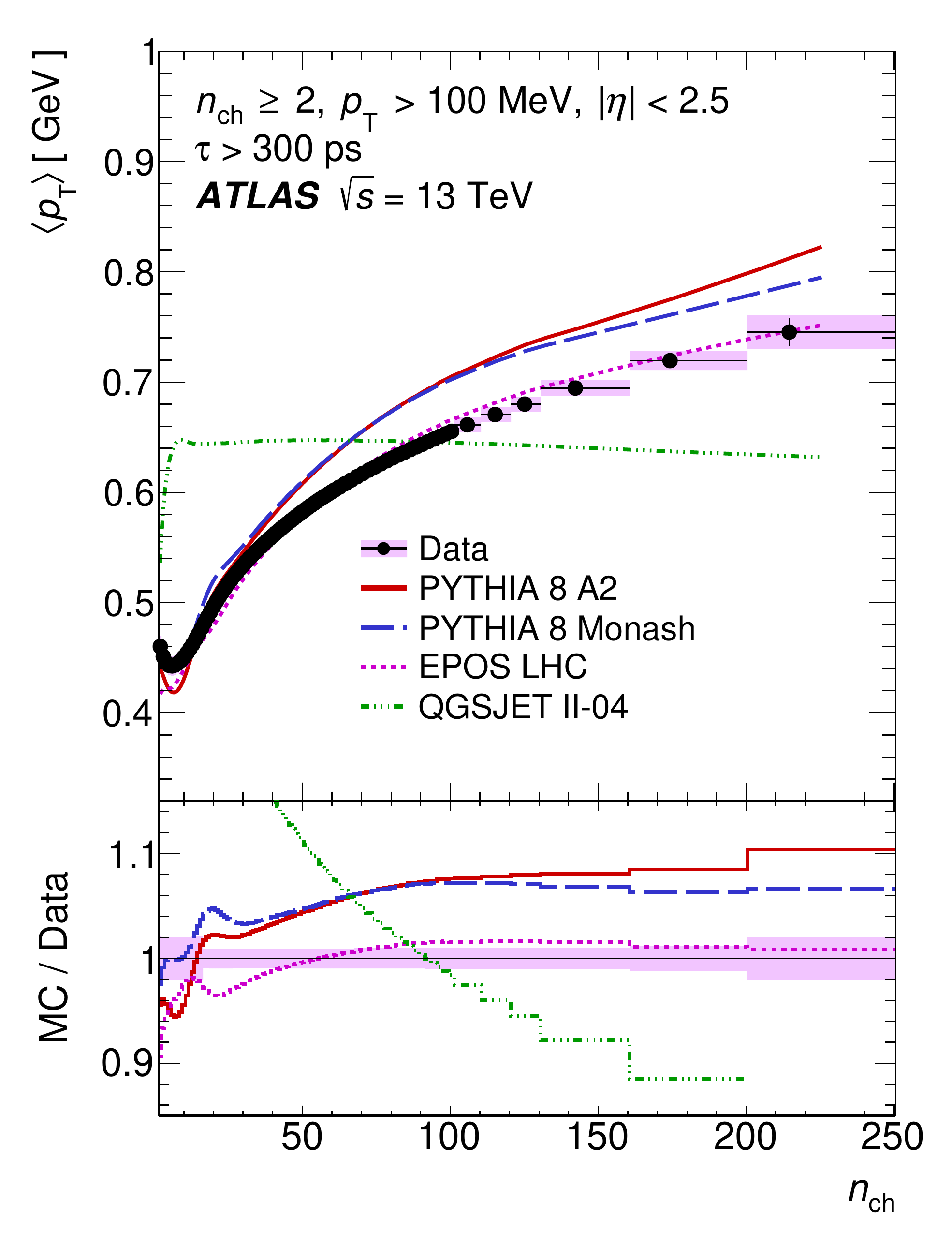}
\label{fig:100MeVptnch}
}
\caption{13 TeV data, from Ref. \cite{MinBias13TeV100MeV}: Primary charged-particle multiplicities as a function of \subref{fig:100MeVEta} pseudorapidity $\eta$ and \subref{fig:100MeVPt} transverse momentum $\pt$,  \subref{fig:100MeVnch} the primary charged-particle multiplicity $\nch$ and \subref{fig:100MeVptnch} the mean transverse momentum $\meanpt$ versus $\nch$ for events with at least two primary charged-particles with $\pt > $ 100 MeV, with $|\eta| <$ 2.5, and with a lifetime $\tau >$ 300 ps. 
}
\label{fig:Final100MeV}
\end{center}
\end{figure}

The corrected distributions for primary charged-particles in events with $\nch =$ 1 in the kinematic range of the {\em nominal phase space} are shown in Figure \ref{fig:Final500MeV}, while, for the {\em extended phase space}, they can be seen in Figure \ref{fig:Final100MeV}. 

Figures \ref{fig:500MeVEta} and \ref{fig:100MeVEta} show the multiplicity of charged-particles as a function of $\eta$. 
In the {\em nominal phase space}, the mean particle density is roughly constant at 2.9 for $|\eta| <$ 1.0 and decreases at higher values of $|\eta|$. EPOS describes the data well for $|\eta| <$ 1.0, and predicts a slightly larger multiplicity at large $|\eta|$ values. 
QGSJET-II and PYTHIA 8-Monash predict multiplicities that are too large by approximately 15\% and 5\% respectively. PYTHIA 8-A2 predicts a multiplicity that is 3\% too low in the central region, but describes the data well in the forward region. The total systematic uncertainty, dominated by the uncertainty on the track reconstruction efficiency, is below 1.5\% in the entire $\eta$ range.
When moving to lower track-$\pt$, the situation changes and PYTHIA 8-Monash, EPOS and QGSJET-II give a good description for $|\eta| <$ 1.5. The prediction from PYTHIA 8-A2 has the same shape as the predictions from the other generators, but lies below the data.
It can be immediately noticed that much bigger systematic uncertainties affect the distribution in the high $\eta$ region in the {\em extended phase space} (up to $\sim$ 7\% with respect to $\sim$ 1.5\% in the {\em nominal phase space}). They mainly come from the uncertainty in the amount of material in the ID, which was differently treated in the two phase spaces, as described above.

Figures \ref{fig:500MeVPt} and \ref{fig:100MeVPt} show the charged-particle transverse momentum distributions. EPOS describes the data well for $\pt >$ 300 MeV. The PYTHIA 8 tunes describe the data reasonably well, but they are below the data in the low-$\pt$ region. QGSJET-II gives a poor prediction over the entire spectrum, overshooting the data in the low-$\pt$ region. 

Figures \ref{fig:500MeVnch} and \ref{fig:100MeVnch}  show the charged-particle multiplicity. PYTHIA 8-A2 describes the data reasonably well in the low-$\nch$ region, but predicts too few events at larger $\nch$ values. PYTHIA 8-Monash, EPOS and  QGSJET-II describe the data reasonably well in the low-$\nch$ region but predict too many events in the mid-$\nch$ region, with PYTHIA 8-Monash and EPOS predicting too few events in the high-$\nch$ region while QGSJET-II, which implements a model without colour-reconnection, describes the data poorly in this case also.

Figures \ref{fig:500MeVptnch} and \ref{fig:100MeVptnch} show how the mean transverse momentum, $\meanpt$, rises versus the charged-particle multiplicity. This rise is expected because of colour coherence effects in dense parton environments and is modelled by a colour reconnection mechanism in PYTHIA 8 or by the hydrodynamical evolution model used in EPOS. EPOS describes the data better than PYTHIA 8, which predicts a steeper rise of $\meanpt$ with $\nch$ than the data. If the high-$\nch$ region is assumed to be dominated by events with a large number of parton interactions within the same pp collision (MPI), without colour coherence effects, the $\meanpt$ should be independent of $\nch$, as predicted by QGSJET-II.

\section{Highlights from 8 TeV measurements}
\begin{figure}[h!]
\begin{center}
\subfigure[]{
\includegraphics[width=0.34\textwidth]{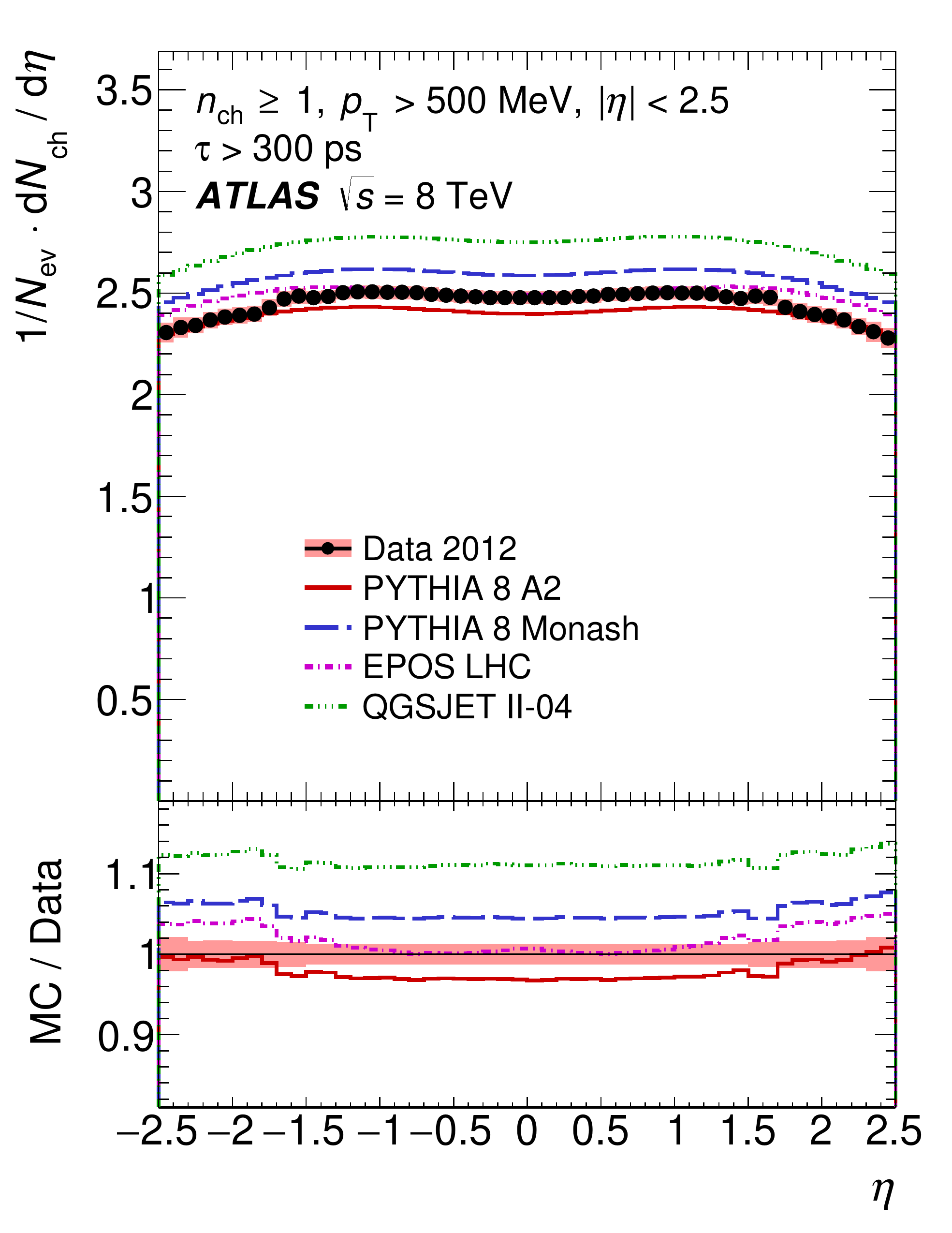}
\label{fig:500MeVEta_8TeV}
}
\subfigure[]{
\includegraphics[width=0.34\textwidth]{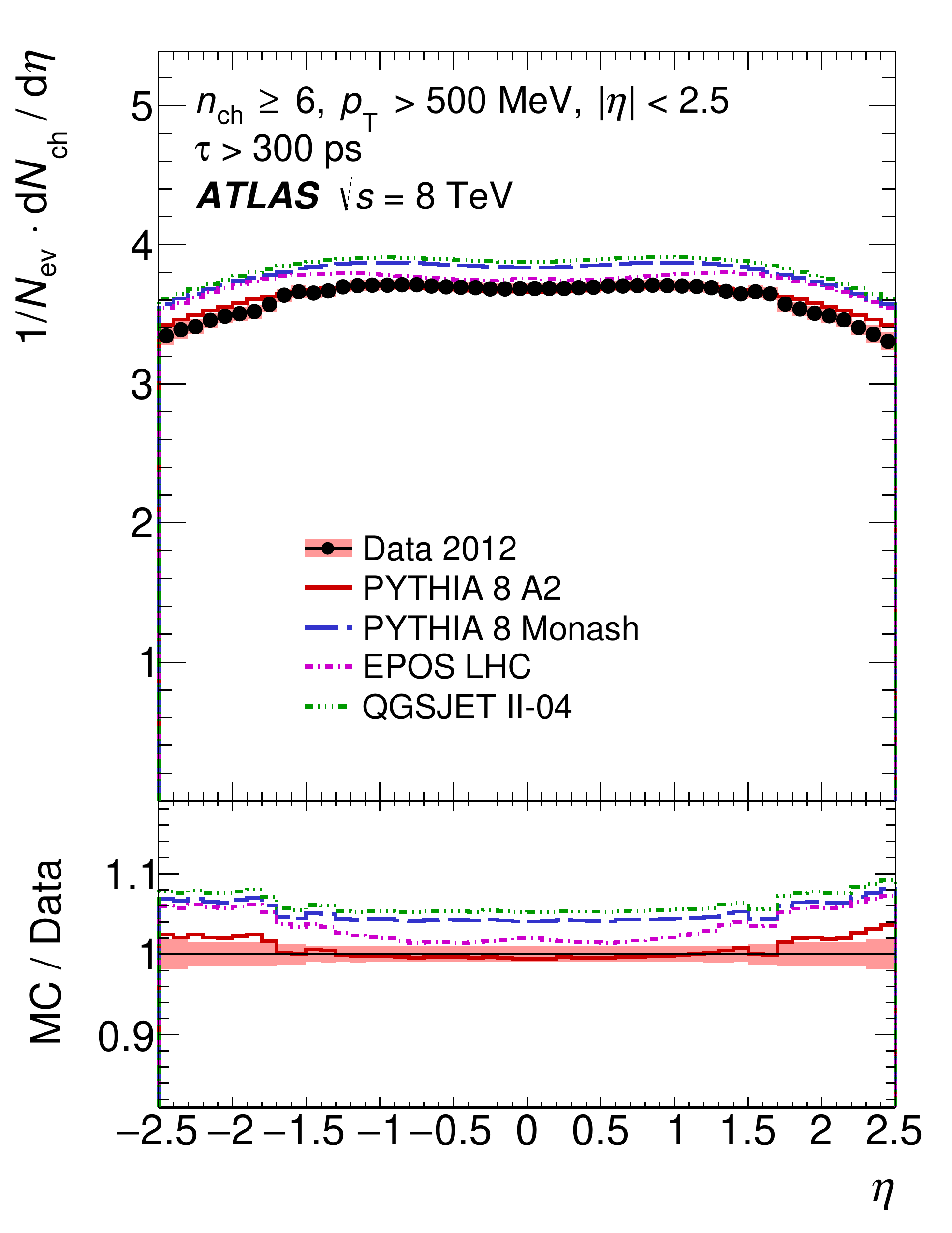}
\label{fig:500MeVEta_8TeV_6}
}
\subfigure[]{
\includegraphics[width=0.34\textwidth]{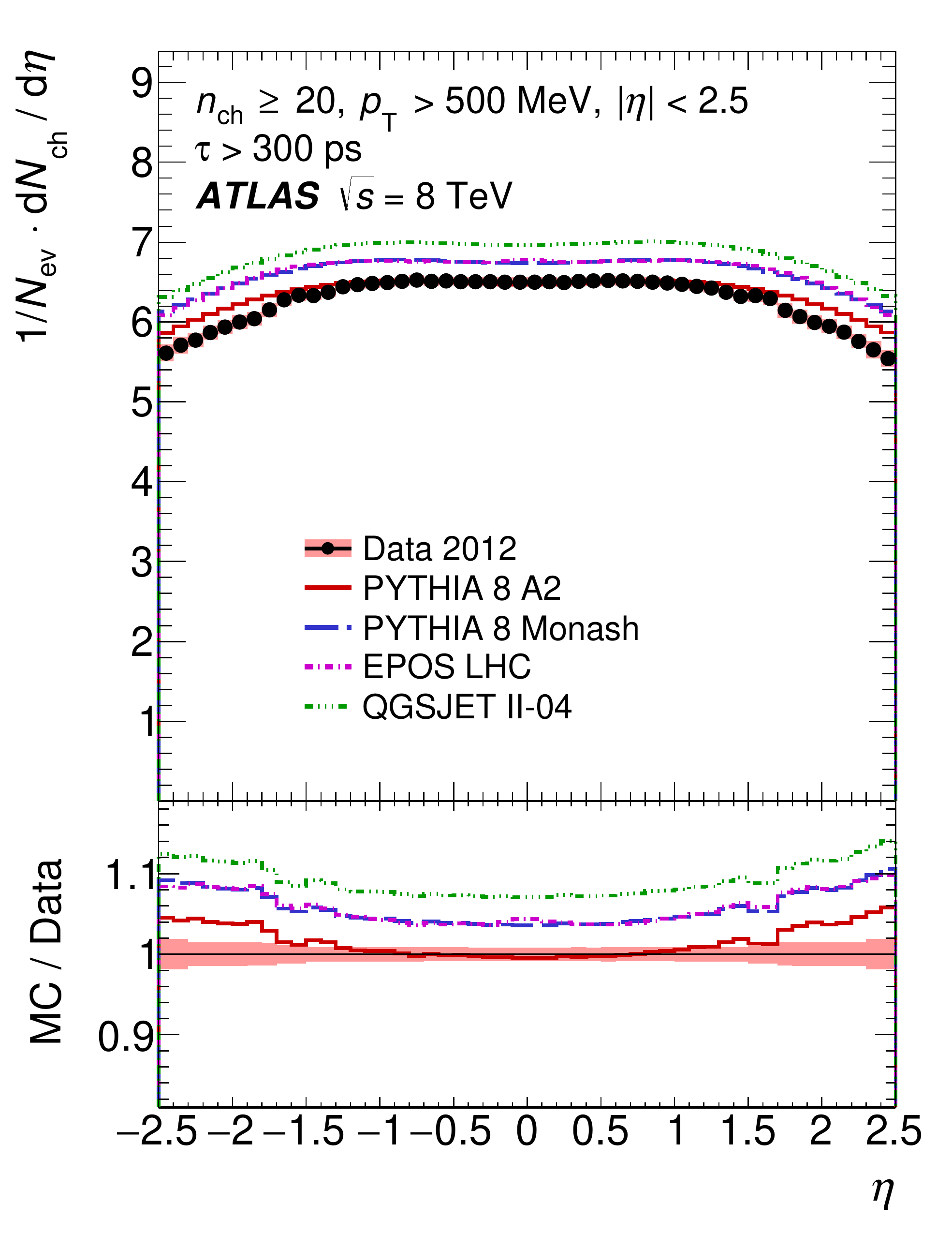}
\label{fig:500MeVEta_8TeV_20}
}
\subfigure[]{
\includegraphics[width=0.34\textwidth]{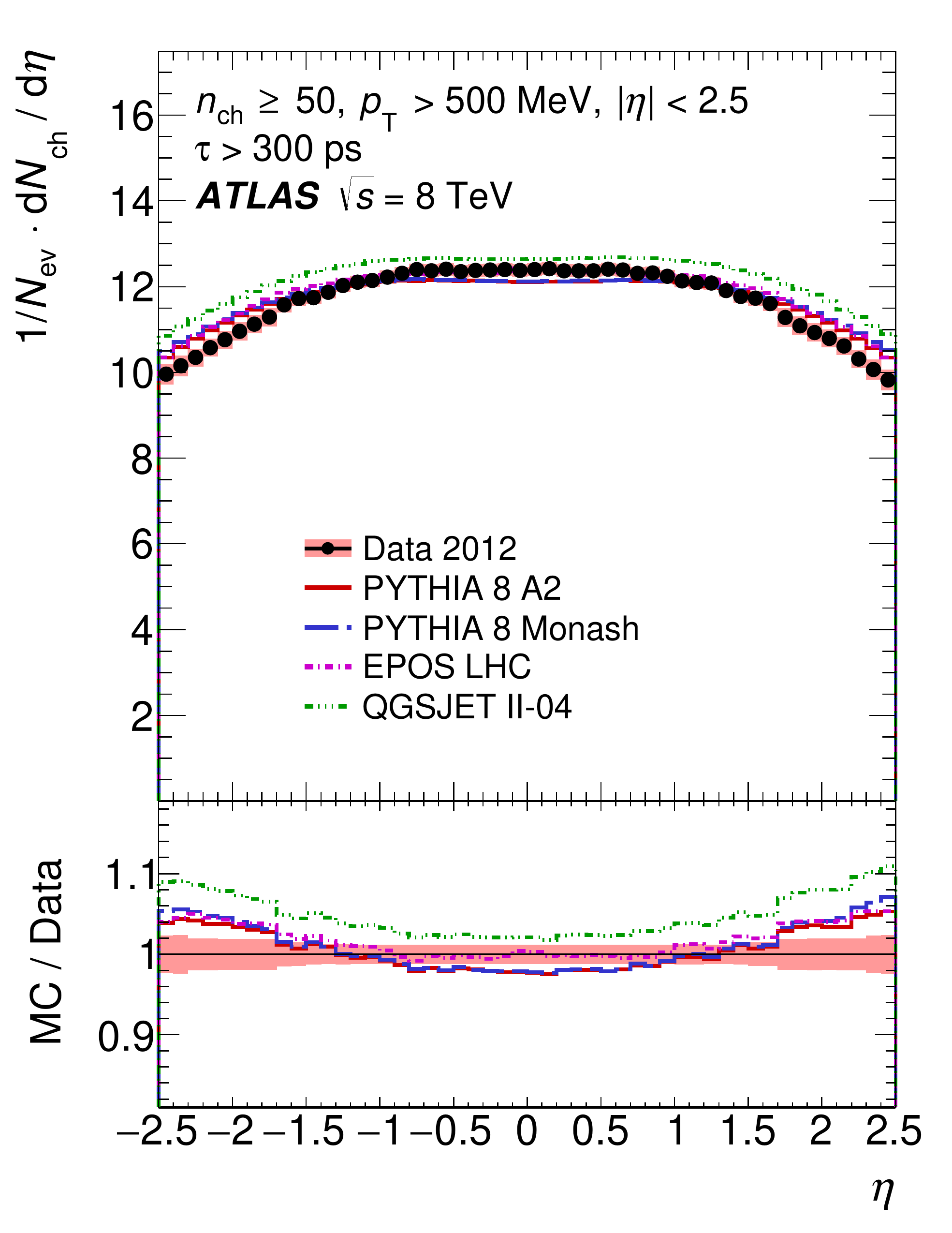}
\label{fig:500MeVEta_8TeV_50}
}

\caption{8 TeV data, from Ref. \cite{MinBias8TeV}: Primary charged-particle multiplicities as a function of pseudorapidity $\eta$ for events with at least \subref{fig:500MeVEta_8TeV} 1, \subref{fig:500MeVEta_8TeV_6} 6, \subref{fig:500MeVEta_8TeV_20} 20 or \subref{fig:500MeVEta_8TeV_50}  50 primary charged-particles with $\pt >$ 500 MeV, with $|\eta| <$ 2.5, and with a lifetime $\tau >$ 300 ps. 
}
\label{fig:Final500MeV_8TeV_highnch}
\end{center}
\end{figure}
\noindent 
In the context of the 8 TeV analysis, high multiplicity phase spaces with $\nch \ge $ 20 and 50 were studied for the first time.
Figure \ref{fig:Final500MeV_8TeV_highnch} shows the multiplicity of charged-particles as a function of $\eta$ for different multiplicity phase spaces with $\nch \ge$ 1, 6, 20, 50. 
When requiring $\nch \ge$ 1, EPOS gives a good description of the data for $|\eta| <$ 1.5, while at higher $|\eta|$, PYTHIA 8-A2 describes the data better.  For $\nch \ge $ 6 or 20, the generator which describes data best is PYTHIA 8-A2. In the $\nch \ge $ 50 case, PYTHIA and EPOS give similar descriptions for $|\eta| >$ 1.5 and overestimate the data, while for $|\eta| <$ 1.5 the best description is given by EPOS.
 
The spectra for the other {\em phase spaces} studied at $\sqrt{s} =$ 8 TeV can be found in \cite{MinBias8TeV}. 

\section{Charged-particle measurements at different $\sqrt{s}$}
\begin{figure}[h!]
\centering
\includegraphics[width=0.44\textwidth]{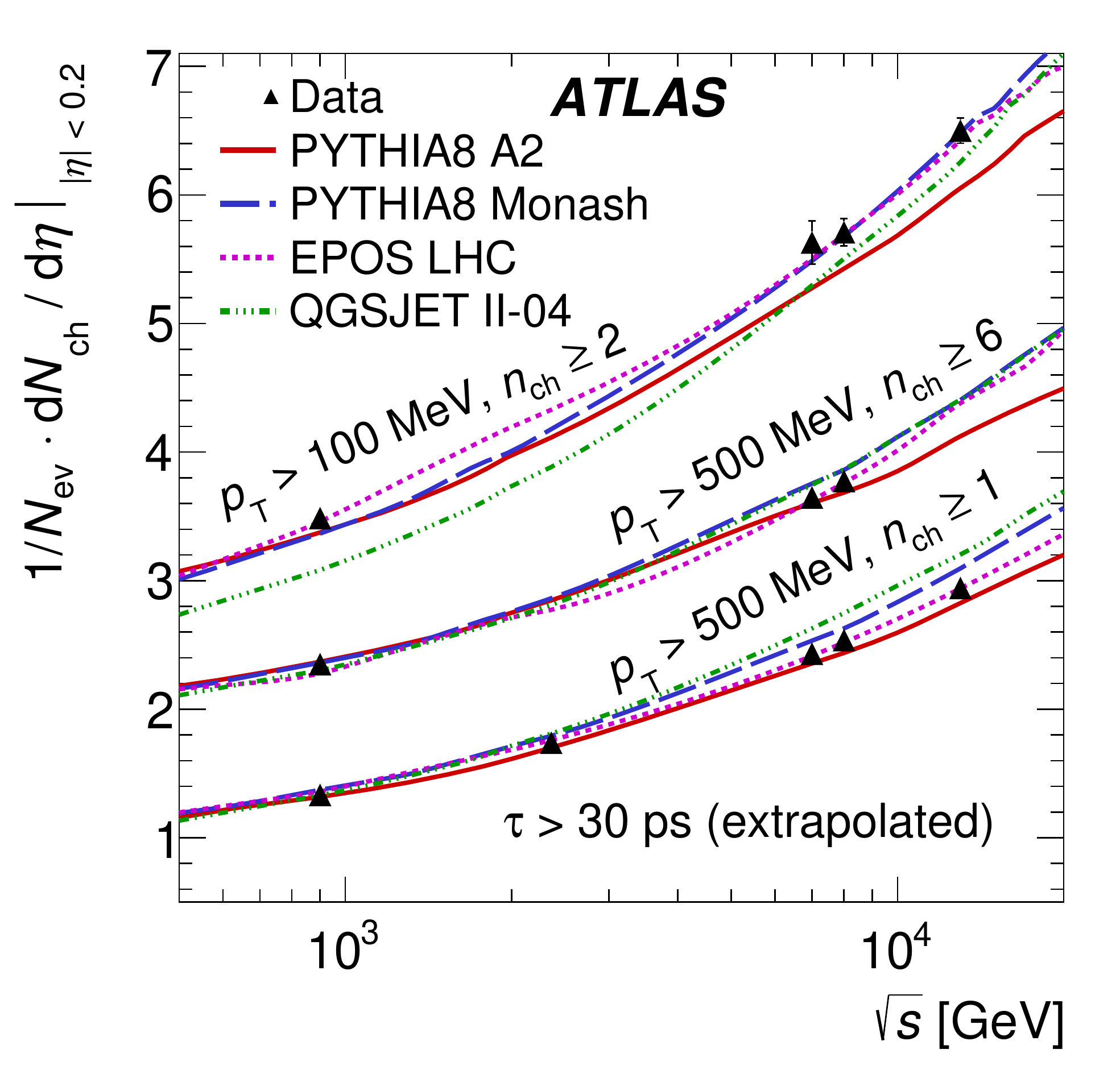}
\caption{The average primary charged-particle multiplicity in $pp$ interactions per unit of pseudorapidity $\eta$ for $|\eta| < 0.2$ as a function of the centre-of-mass energy $\sqrt{s}$, from Ref. \cite{MinBias13TeV100MeV}. The results at 8 and 13 TeV have been extrapolated to include charged strange baryons in order to compare the values with previous studies. 
}
\label{fig:MinBiasFinalPlot}
\end{figure}
\noindent 
Figure \ref{fig:MinBiasFinalPlot} shows the mean number of primary charged-particles in the central region of the detector as a function of $\sqrt{s}$. It is obtained by averaging over $|\eta| <$ 0.2 and by correcting the 8 and 13 TeV data for the contribution from strange baryons in order to compare the results with lower centre-of-mass energies at which these particles were included.
The mean number of primary charged-particles increases by a factor of 2.2 when $\sqrt{s}$ increases by a factor of about 14 from 0.9 TeV to 13 TeV. EPOS describes the dependence on $\sqrt{s}$ very well in several phase spaces, while PYTHIA 8-A2 and PYTHIA 8-Monash give a reasonable description, respectively for $\pt >$ 500 MeV and $\pt >$ 100 MeV. QGSJET-II predicts a steeper rise in multiplicity with $\sqrt{s}$ than that shown by the data.

\section{Conclusions}
Primary charged-particle multiplicity measurements with the ATLAS detector using proton-proton collisions delivered by the LHC at the centre-of-mass energy of 13 TeV are presented, with a particular emphasis on the tracking-related aspects. Some highlights from the high charged-particle multiplicity regions studied at the 8 TeV centre-of-mass energy are also given.
The results show clear differences between MC models and the measured distributions. Among the models considered, EPOS reproduces best the data, PYTHIA 8-A2 and PYTHIA 8-Monash give reasonable descriptions of the data and QGSJET-II fails to describe most of the features of the data.


\end{document}